\documentclass[english]{article}
\usepackage{geometry}
\usepackage{amsmath}
\usepackage{amssymb}
\usepackage{graphicx}
\usepackage{setspace}
\usepackage{epstopdf}
\usepackage{authblk}
\usepackage[authoryear]{natbib}

\makeatletter

\providecommand{\tabularnewline}{\\}

\setcitestyle{round}
\usepackage{babel}
\date{}

\makeatother

\usepackage{babel}
\begin{document}

\title{A Bootstrap Method for Goodness of Fit and Model Selection\\with a Single Observed Network}
\author[1]{Sixing Chen}
\author[1]{Jukka-Pekka Onnela}
\affil[1]{Department of Biostatistics, Harvard T.H. Chan School of Public Health, Boston, MA}
\affil[ ]{sixingchen@hsph.harvard.edu, onnela@hsph.harvard.edu}

\maketitle
\begin{abstract}
Network models are applied in numerous domains where data can be represented
as a system of interactions among pairs of actors. While both statistical
and mechanistic network models are increasingly capable of capturing
various dependencies amongst these actors, these dependencies
imply the lack of independence. This poses statistical challenges
for analyzing such data, especially when there is only a single observed
network, and often leads to intractable likelihoods regardless of
the modeling paradigm, which limit the application of existing statistical
methods for networks. We explore a subsampling bootstrap procedure
to serve as the basis for goodness of fit and model selection with
a single observed network that circumvents the intractability of such
likelihoods. Our approach is based on flexible resampling distributions
formed from the single observed network, allowing for finer and higher
dimensional comparisons than simply point estimates of quantities
of interest. We include worked examples for model selection, with
simulation, and assessment of goodness of fit, with duplication-divergence
model fits for yeast (\textit{S.cerevisiae}) protein-protein interaction
data from the literature. The proposed procedure produces a flexible
resampling distribution that can be based on any statistics of one's
choosing and can be employed regardless of choice of model.
\\
\\
\noindent \textbf{Keywords} single empirical network, network models, resampling, model selection, goodness of fit
\end{abstract}

\newpage{}

\section{Intro}

Networks are well-suited to represent the structure of data from systems
composed of interactions between pairs of actors (represented by nodes)
that make up the system \citep{mej2010networks,wasserman1994social,pastor2007evolution,lusher2013exponential,NetworkExpRaval2013introduction}.
Often in such systems, these interactions (represented by edges) can
depend on the state of the rest of the system, such as existing edges
as well as attributes of nodes. One prominent example of this
is triadic closure in social networks, where two people are more likely
to become friends should they share a mutual friend \citep{watts2004six}. While innovations
in network models are increasing the capability to encompass various
dependencies between edges in the data, this rich level of interconnectedness
poses a problem for statistical methods for networks.

In typical statistical settings, the premise is that the data is
composed of independent observations. Typical methods are able to
derive efficiency gains and consistency from a large number of samples
due to this independence. However, in the network context where the
structure of the network is of primary interest, the edges and their placement can
be seen as the outcome, but there are often multiple layers of between-edge
dependence. Thus, the premise of independent observations may not be met
and most available statistical methods are therefore not applicable.

To see how limited statistical methods are for networks, one can
inspect two prominent paradigms of network models.
Statistical models are probabilistic models that specify the likelihood
of observing any given network \citep{robins2007introduction,LatSpaceHoff2002latent,goyal2014sampling}.
One example is the family of exponential random graph models (ERGMs)
\citep{lusher2013exponential}, which use observable configurations
(such as triangles and $k$-stars) as the natural sufficient statistics.
Although popular in practice, ERGMs can be difficult to fit and to
sample from, and related methods may not scale well with large networks
\citep{an2016fitting}. Estimation for ERGMs can proceed via maximum
pseudolikelihood estimation (MPLE) \citep{besag1974spatial} or Markov
chain Monte Carlo maximum likelihood estimation (MCMC-MLE) \citep{geyer1992constrained,snijders2002markov}.
Pseudolikelihood methods for inference in ERGMs can lead to biased
results due to the ignored dependence \citep{van2009framework}, while
inference for MCMC-MLE proceeds via simulation from estimated model
\citep{snijders2002markov}, and is thus entirely model based. On
the other hand, mechanistic models are composed of generative mechanisms
that prescribe the growth and evolution of a network over time \citep{barabasi1999emergence,watts1998collective,sole2002model,vazquez2003modeling,klemm2002highly,kumpula2007emergence}.
While they are easy to sample from, a mechanistic model allows for
numerous paths that can be taken in the state space to produce any one observed network,
making the likelihood of all but the most trivial models intractable
for networks of modest size. As a result, performing statistical procedures
is difficult for such models and have little extant work in the literature.

In situations where likelihood based methods are not available, one
often resorts to resampling methods, such as bootstrap, jackknife,
and permutation tests \citep{efron1981nonparametric,good2006resampling,wu1986jackknife}.
Although the different resampling methods operate differently, they
all serve to create new data sets from a single observed data set
that mimic the behavior of the original one to serve as a basis for
statistical procedures. This can be an attractive option for networks,
since the data set can often consist of a single observed network.
Examples include the internet and the world wide web, large social networks,
certain biological networks, as well as transportation and infrastructural networks, to name a few.
Having multiple resampled networks that resemble, in some ways, the
original observed network can allow one to bypass dealing with the
unwieldy likelihoods of current network models. Even in the best case,
despite the likelihood having a simple functional form, the normalizing
constants of ERGMs are generally unobtainable, since they require
summing over an astronomical number of possible network realizations
even for a network of modest size. In this paper, we
will explore using a resampling procedure as a basis for statistical
procedures for a single observed network.

There is some existing research on resampling methods in settings
involving networks. First, there are methods for assessing the goodness
of fit for a fitted model \citep{hunter2008goodness,shore2015spectral}.
These methods generally work by drawing network realizations from
the fitted model, then assessing fit by comparing the value of a set
of statistics for the observed network to the distribution of said
statistics of the generated draws. This resampling scheme is akin
to that of the parametric bootstrap. Note that this can be done for
the point estimate of individual statistics or those of multiple statistics
simultaneously, e.g., functionals of the degree distribution. However, the resamples
in these methods are only representative of the fitted model and not
necessarily of the observed network, and comparisons are made
based only on point estimates. Second, there are methods for a setting
where there are multiple independent networks observed for MPLE \citep{desmarais2012statistical}.
This is similar to the typical statistical setting with multiple independent
observations and not applicable to the setting with just one observed
network. Lastly, there are resampling methods based on subgraphs of
subsamples of nodes in the observed network \citep{ohara2014resampling,bhattacharyya2015subsampling,ali2016comparison,thompson2016using,gel2017bootstrap}.
\citet{ohara2014resampling}, \citet{bhattacharyya2015subsampling}, \citet{thompson2016using},
and \citet{gel2017bootstrap} are aimed at estimation and uncertainty quantification
of network centrality, distribution of subgraphs, and functionals of the degree
distribution, while \citet{ali2016comparison} is a subgraph-based method for
comparison between networks.

The procedure we propose makes use of the bootstrap subsampling scheme
from \citet{bhattacharyya2015subsampling}. Our proposed
boostrap method addresses goodness of fit and model selection rather
than estimation, and is based on the resampling distribution (rather than
point estimates) of any set of statistics obtained from the induced subgraphs.
The flexible choice of statistics allows an investigator
to focus the criterion for model fit based on the aspects of the network of
scientific interest. The flexibility of the full resampling distribution contains more
information than simply aggregated subgraph counts and point
estimates for comparison with candidate models. It also allows for
natural uncertainty quantification regardless of the algorithm used for
selecting the model. The proposed procedure is agnostic
to the modeling paradigm (statistical or mechanistic) and can accommodate
any model from which one can sample from, while providing very interpretable results.
The scaling of the procedure depends on that of the statistics chosen as well
as the number of subsamples taken. The latter is the only component native
to our procedure and is linear.

The rest of the paper is organized as the following. In sections 2
through 4, we explain the proposed bootstrap subsampling procedure
in detail, and highlight important considerations for some of the steps.
In section 5, we elaborate on potential scenarios for when the proposed
procedure could be used, and some of these are showcased with simulations and data
example in section 6. Lastly, we conclude with discussions in section
7.

\section{Subsampling Scheme and Resampling Distributions}

Each subsample of the bootstrap subsampling scheme of \citet{bhattacharyya2015subsampling}
consists of a uniform node-wise subsample of all the nodes in the
observed network $G_{o}$ (with node set $V_{o}$ and edge set $E_{o}$)
and their induced subgraph, i.e., the nodes in the subsample and all
edges between these nodes. For each subsample, one may compute any
set of statistics to form a resampling distribution of these statistics.
Although the subsamples will not be representative of a network
the same size as the subsample from the true data generating mechanism,
they will still retain features of the true data generating mechanism
since the subsampling does not directly change any between-edge and
between-node dependence that influenced the formation of the network,
despite adding a degree of ``missingness'' by removing elements
correlated with those in the subsample. In comparison, should one
generate draws from a particular fitted model in order to form a resampling
distribution, then the between-edge and between-node dependence will
be those specified by the fitted model. In this case, the generated
networks will only be representative of the true data generating mechanism
if the fitted model is the true model, which is a strong assumption
in most cases, and usually not verifiable in practice.

Due to each subsample only consisting of a subsample of $V_{o}$ and
$E_{o}$, each subsample will be missing elements that are correlated
with those that are included in the subsample. As a result, this must
be taken into account when any comparisons are made with a null/candidate
model $M_{c}$. One may be tempted to compare subsamples of $G_{o}$
with draws from $M_{c}$ of the same size as the subsample. This should
however be avoided since there is a degree of ``missingness'' in
the subsamples of $G_{o}$ that are not present in such draws from
$M_{c}$. Even if $M_{c}$ was the true model, this disparity could
make the two behave differently. Instead, one should generate draws
from $M_{c}$ the same size as $G_{o}$ and then apply the same subsampling
to these draws. This way, both the subsamples of $G_{o}$ and those
of $M_{c}$ will display the same amount of ``missingness''
and will be comparable. Should $M_{c}$ be representative of the true data
generating mechanism, then behavior of the two subsamples and the
resampling distributions of computed statistics should be similar.
The representativeness of the subsamples from $G_{o}$, as well as
this comparability with the subsamples from $M_{c}$, form the basis
for our statistical procedures. Even though we only consider uniform subsampling
in the paper, the method for subsampling is flexible and can
be chosen so that it is representative of sampling in practice or
for statistical and computational ease. The proposed bootstrap subsampling
procedure is summarized in \textbf{Figure} 1.

\begin{center}
\includegraphics[width=\linewidth]{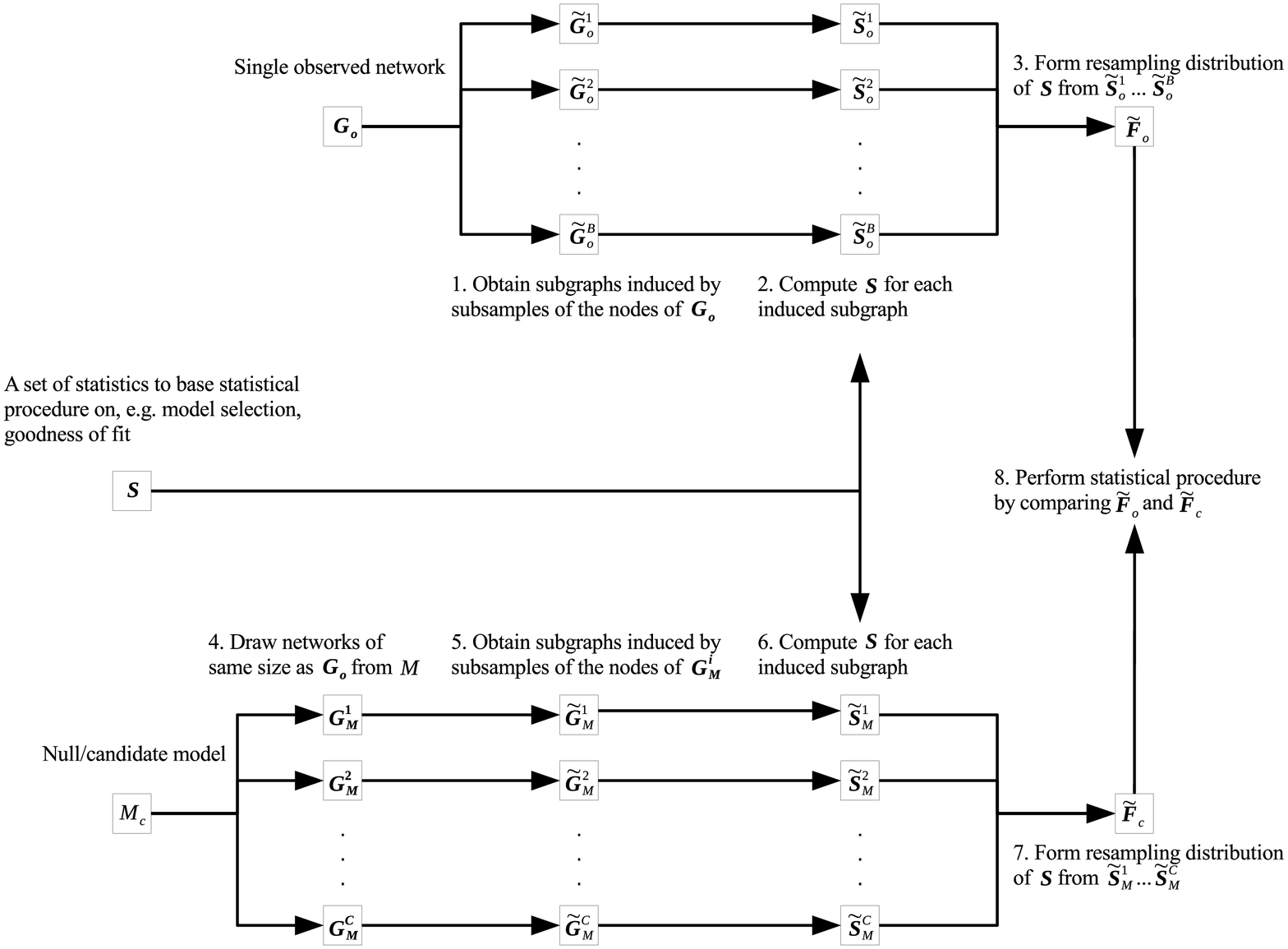}
\par\end{center}
\begin{description}
\item [{Figure}] 1: Schematic of the steps of the proposed bootstrap subsampling
procedure for a single observed network $G_{o}$.
\end{description}

In contrast to existing methods that also use draws from the fitted
model to assess goodness of fit, this approach can lead to a richer
level of comparison. For exisiting methods, after choosing the statistics
desired for assessing goodness of fit, the given statistics are computed
for $G_{o}$ and a large number of draws from $M_{c}$. The point
estimate of these statistics for $G_{o}$ are placed within the distribution
of said statistics of the draws from $M_{c}$. Goodness of fit is
then assess by the location of the point estimate from $G_{o}$ within
the draws from $M_{c}$. This can be done visually or by quantifying
the proportion of the draws with values of the statistics deemed more
extreme. With our approach, the two resampling distributions can be
compared on multiple levels, such as their location, spread, and shape.
In addition, one can quantify the distance between the two with statistics
such as the Kolmogorov-Smirnov (KS) statistic or the Kullback-Leibler
divergence to order the fit of different candidate models.

One point of interest is that the subsamples from $G_{o}$ are all
from a single network, while subsamples from $M_{c}$ are subsamples
of independent networks drawn from $M_{c}$ instead of subsamples
from a single network drawn from $M_{c}$. The former is proposed
due to potential instability of single generated networks and the
corresponding subsamples, since there can be a great deal of instability
in the generated networks depending on the model and the seed network
used (often required to grow networks specified by mechanistic models).
In addition, the disparity between the two styles of subsamples
may depend on the proportion of the nodes in each subsample. Both of these
points are further examined in the next two sections.

\section{Stability under Sampling}

When sampling from the candidate model, one needs to take care so
the draws actually behave like the observed network even if the candidate
model is the true model or is an accurate model, and in turn, the
subsamples of these draws behave like the subsamples of the observed
network. Such draws can look nothing like the observed network despite
having a good candidate model, e.g., the draws having highly varying
degree distributions that look nothing like that of the observed network.
This issue can be more prominently demonstrated in the context of
some mechanistic network models.

Networks generated from mechanistic models are often grown from a
small (relative to the final size of the network) seed network according
to the model's generative mechanism until some stopping condition
is reached, e.g. attaining a requisite number of nodes. There has
works that show the original seed network has no influence on the
degree distribution in the limit, i.e., for a large number of nodes, for
certain types of mechanistic network models \citep{cooper2003general,li2013degree}.
While some data sets, such as social networks, may be sufficiently
large to reach this asymptotic regime, others, such as protein-protein
interaction networks, may not be. Thus, when generating draws from
candidate models for analysis of smaller networks, the original seed
network can potentially have a great deal of influence. The seed network
maybe as simple as a single node, or a complete graph of only three
nodes, up to bigger complete graphs, or something more elaborate with
more than one component. We briefly examine the effect of the seed
network on the stability of the degree distribution of networks generated
from the Erd\H{o}s-R\`enyi and duplication-divergence models, of
protein-protein interaction networks.

\subsection{Erd\H{o}s-R\`enyi Model}

The Erd\H{o}s-R\`enyi (ER) model \citep{ERerdds1959random}
is a simple but rather unique model in that it can be framed as both
a mechanistic and a statistical model. In the ER model, the number
of nodes $n$ is fixed, but there are two variants of the model that
determine how the edges are placed. In the first variant, the $G(n,p)$
model, each of the $C\left(n,2\right)$, $n$ choose 2, possible edges
are independent and are included in the graph with probability $p$,
so the number of edges in the graph is binomial. In the other variant,
the $G(n,m)$ model, the number of edges in the graph $m$ is also
fixed. In this case, the random graph has a uniform distribution over
all $C\left(C\left(n,2\right),m\right)$ possible graphs with $n$
nodes and $m$ edges.

\begin{center}
\includegraphics[width=\linewidth]{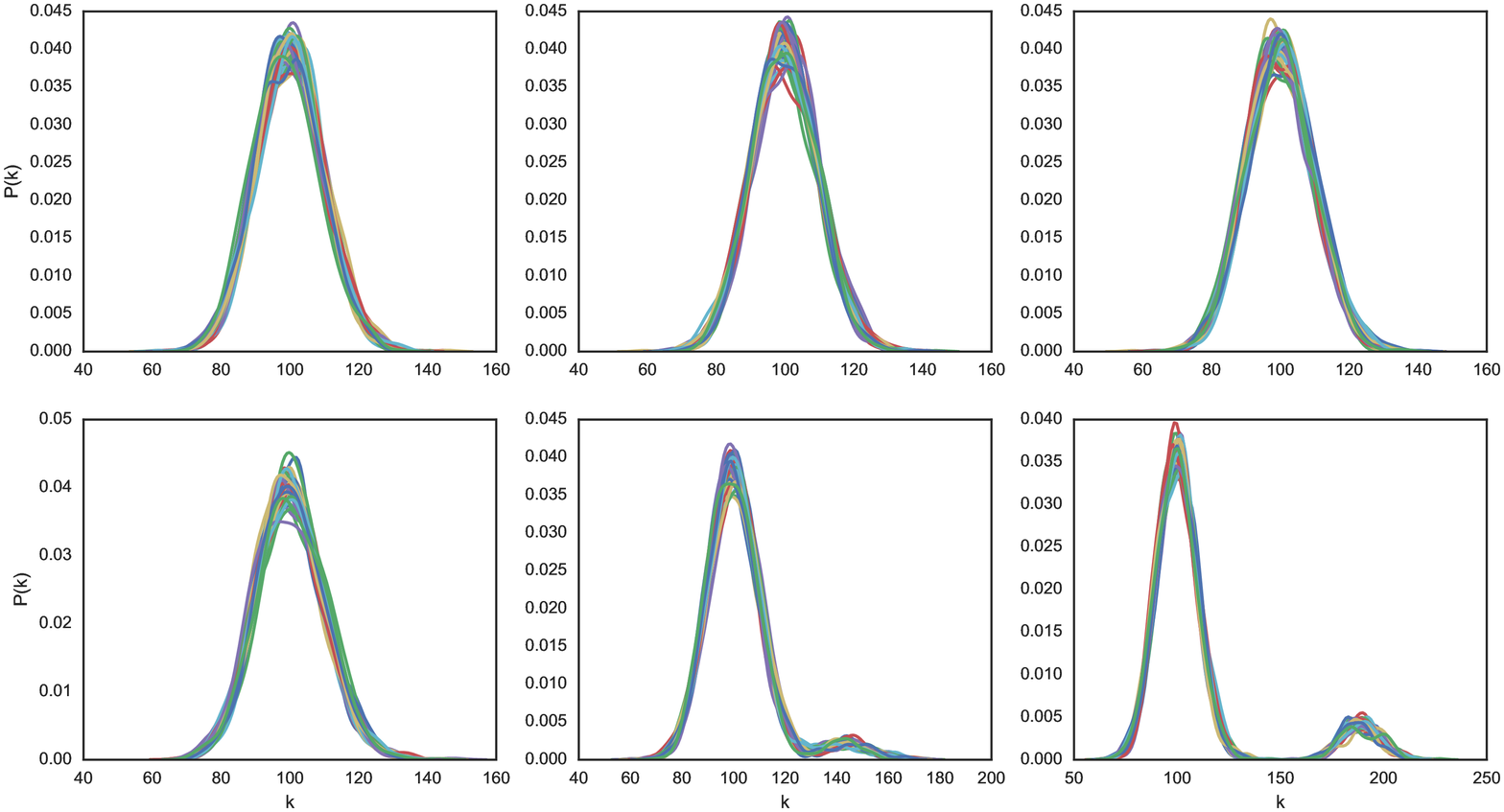}
\par\end{center}
\begin{description}
\item [{Figure}] 2: The degree distribution of 50 generated graphs from
the $G\left(n=1000,p=0.1\right)$ model with seeds of 5, 8, 10, 20,
50, 100 nodes, from left to right, then top to bottom, as described in text.
\end{description}

The first variant can be easily framed as a mechanistic model. The
network generation starts with a seed network of a single node. Then
at each stage, a new node is added, and an edge between the new node
and each existing node is added with probability $p$. This is done
until there are $n$ nodes in the network. Rather than starting with
a seed network of a single node, networks can be generated according
to the generative mechanism of the $G\left(n,p\right)$ model initialized
with a different seed network. Here, we generated $G\left(n=1000,p=0.1\right)$
networks according to these rules, with complete
graphs of 5, 8, 10, 20, 50, 100 nodes as the seed networks. We generated
50 networks of each size of the seed to evaluate the influence of
the seed network on the stability of the degree distribution of the
fully grown network.

The degree distribution of the 50 generated graphs at each size of
the seed network are plotted in \textbf{Figure} 2. While the shape
of the degree distribution understandably changes as the complete
graph used as the seed network gets bigger, the size of the seed network
seems to have little influence on the stability of the degree distribution.
All 50 networks, for each size of the seed network, have very similar
degree distributions. The width of the ``band'' of the 50 distributions
stacked on top of one another also looks to be mostly unchanging.
This seems to indicate that the variability in the degree distribution
is largely unaffected by the size of the seed network.

\subsection{Duplication-Divergence Models}

Duplication-divergence models are a popular class of models used for
protein-protein interaction networks. Examples include the duplication-mutation-complementation
(DMC) \citep{vazquez2003modeling} and duplication-mutation-random
mutation models (DMR) \citep{sole2002model,pastor2003evolving}. Given
a seed network, both DMC and DMR models grow the network according
to their respective generative mechanisms until the requisite number
of nodes, $n$, is reached. In both the DMC and DMR models, a new
node is first added at the beginning of each time step in network generation.
An existing node is chosen uniformly at random for duplication, and an edge is then
added between the new node and each neighbor of the chosen node. After
this, the two models diverge. For DMC, for each neighbor of the chosen
node, either the edge between the chosen node and the neighbor or
the edge between the new node and the neighbor is removed with probability
$q_{mod}$. The step is concluded by adding an edge between the chosen
node and the new node with probability $q_{con}$. For DMR, each edge
connected to the new node is removed with probability $q_{del}$.
The step concludes by adding an edge between the new node and any
existing node at the start of time step $t$ with probability $q_{new}/n\left(t\right)$,
where $n\left(t\right)$ is the number of nodes in the network at the start of time step $t$.

To assess stability of the degree distribution, we generated 50 network
realizations of 1000, 3000, 5000, 7000, 10000 nodes from both models
with the seed network set as a complete graph with 5, 8, 10, 20, 50,
100 nodes. The parameters of the DMC model were set as $q_{mod}=0.2$
and $q_{con}=0.1$, while those of the DMR model were $q_{del}=0.2$
and $q_{new}=0.1$. The degree distribution for the 50 generated networks
at each combination of the size of the seed network and the total
number of nodes for both models are plotted in \textbf{Figures} 3
and 4. A general trend in the plots is that the total number of nodes
in the network has little to no influence on the stability of the
degree distribution, while the size of the seed network has a great
deal of influence, with stability increasing sharply with the size
of the seed network, up to 50. For smaller seed networks, i.e., 3 or
5, the shape and spread of the degree distributions vary wildly even
for larger networks. With a modest increase in the size of the seed
network, i.e., 8 or 10, the shape and the spread of the degree distributions
are more similar. Finally, for larger seed networks, i.e., 20, 50, or
100, the shape and spread of the degree distributions are quite uniform,
and the width of the ``band'' of the 50 degree distributions stacked
on top of one another also decreases. Clearly, the variability of
the degree distribution depends greatly on the size of the seed network.

\begin{center}
.\includegraphics[width=\linewidth]{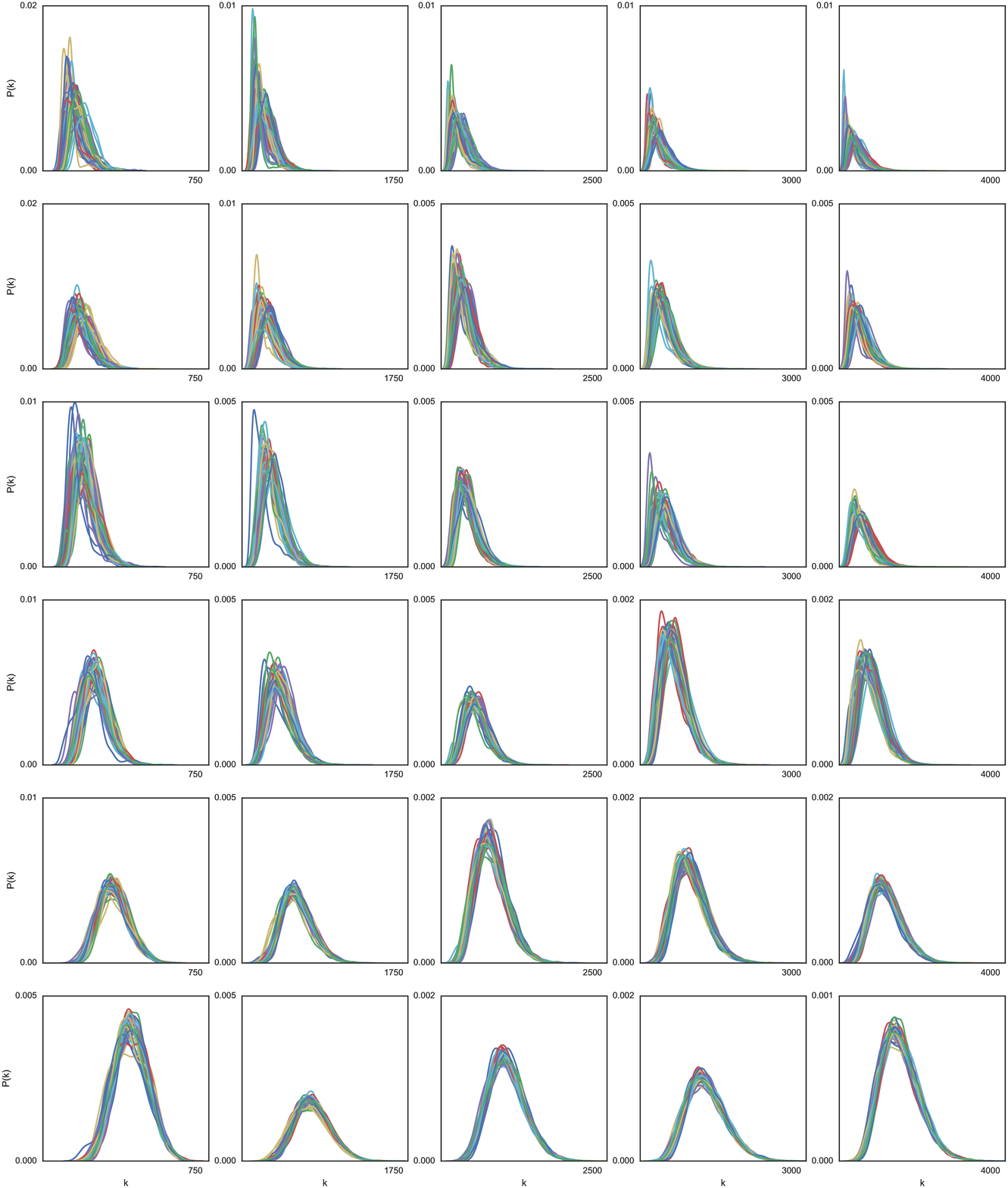}
\par\end{center}
\begin{description}
\item [{Figure}] 3: The degree distribution of 50 generated graphs of
1000, 3000, 5000, 7000, 10000 nodes from the DMC model, from left
to right, with seeds of 5, 8, 10, 20, 50, 100 nodes, from top to bottom.
\end{description}

\begin{center}
\includegraphics[width=\linewidth]{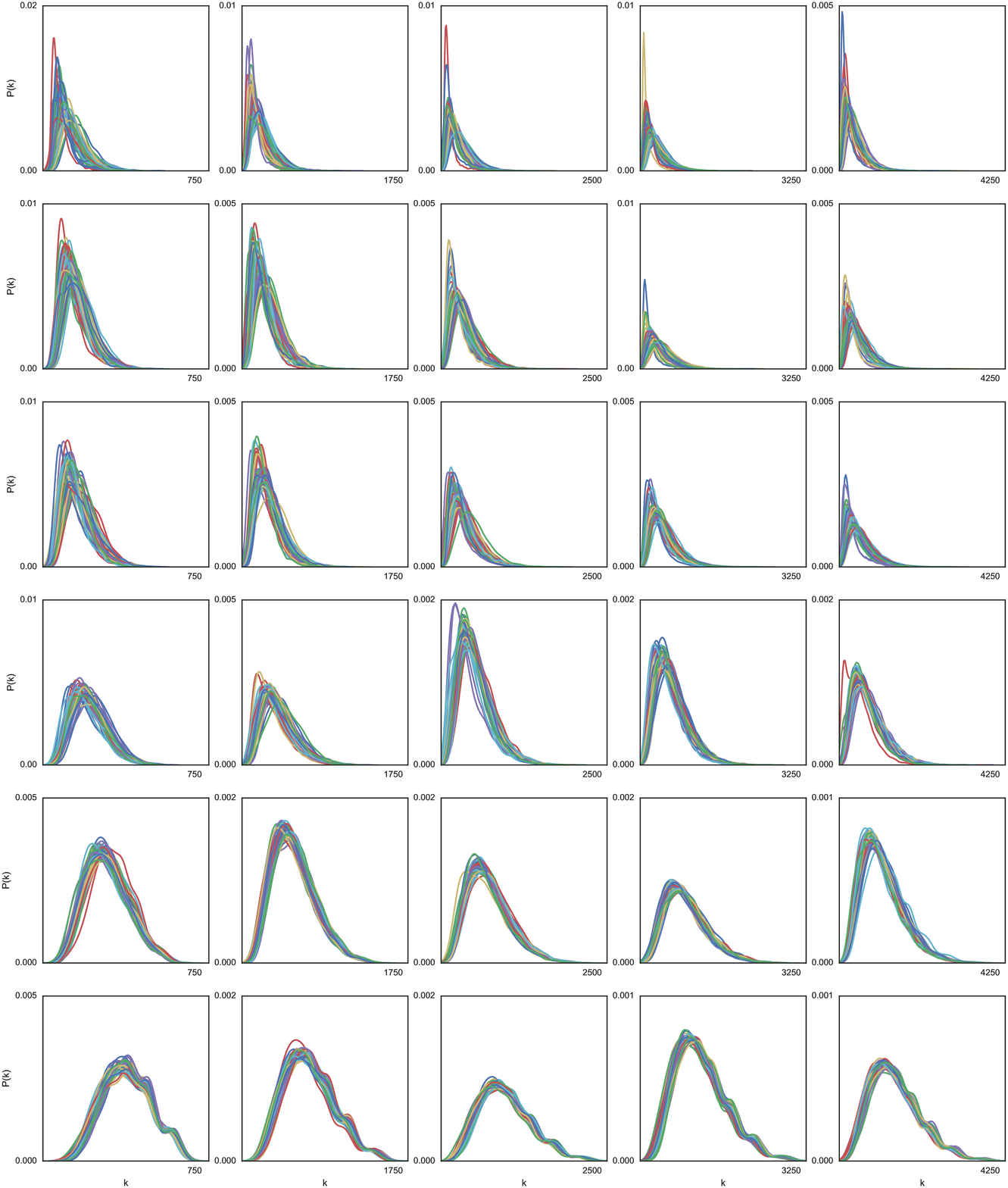}
\par\end{center}
\begin{description}
\item [{Figure}] 4: The degree distribution of 50 generated graphs of
1000, 3000, 5000, 7000, 10000 nodes from the DMR model, from left
to right, with seeds of 5, 8, 10, 20, 50, 100 nodes, from top to bottom.
\end{description}
\newpage{}

One big difference between the ER and DMC/DMR models
is the dependence on exisiting edges on the formation of new ones.
The instability in the degree distribution of networks generated from
DMC/DMR models with small seed networks can be
attributed to this dependence. While these two examples show the influence
the seed network can potentially have in generating networks of modest
size with mechanistic models, it does beg the question of how one
selects a meaningful seed that leads to stable sampling while mimicking
the behavior of the observed network in a principled way. Hypothetically,
if the observed network is indeed generated from an ER model and assuming
the seed network and the parameter values are well chosen, then the
generated networks should mostly appear similar to the observed network
due to the low variability regardless of the size of the seed. On
the other hand, should the observed network come from a DMC/DMR
model and assuming well chosen parameter values, as well as an appropriate
but small seed network, then the generated networks are unlikely to
appear similar to the observed network due to the high variability
with small seeds as demonstrated.

\section{Portion of Nodes to Include in Subsamples}

The portion of nodes included in each subsample should not be so small
such that no characteristics of the observed network or candidate
models are retained, but also not so big such that the subsamples
contain little variability. In one extreme, each subsample consists
of just one node so that there is no structure within the induced
subgraph, and in the other extreme, each subsample is simply the entire network.
While the latter is of little concern when taking subsamples from
independent draws from candidate models, it leaves no variability
in the subsamples from a single observed network such that any resulting
resampling distribution would simply be a point mass. What is an appropriate
portion of nodes to include in each subsample?

Before attempting to answer this question, we define a criterion for
performance in terms of the expectation of the KS statistic (lower values
are better) between $F_{1}$, the resampling distribution from the
subsamples of a single network drawn from candidate model $M_{c}$,
and $F_{c}$, that from subsamples of several independent networks drawn from
$M_{c}$, where each subsample comes from different independent draws.
This quantity is a measure of how closely $F_{o}$, the
resampling distribution from the subsamples of the observed network,
match $F_{c}$ when the observed network is truly generated by $M_{c}$.
If the KS statistic is small, discrepency between $F_{o}$ and $F_{c}$
will be small if the model is correct. Additionally, this quanity being
small implies that there is not much difference between using $F_{1}$
and $F_{c}$ for comparison with $F_{o}$, thus we would be better
off in electing for the stability of $F_{c}$. Note that the computation
time required for $F_{c}$ is greater than that for $F_{1}$.

To compute the expectation of this KS statistic in general is not possible, since it largely
depends on the network model and the seed network used. We will examine this
quantity in the setting of the above mentioned $G\left(n,p\right)$
variant of the ER model, where the resampling distribution is the
edge count in the induced subgraphs. We chose this model since the induced
subgraph of an ER graph is once again an ER graph, so the distribution
of the number of edges is still binomial and tractable.

The desired expectation of the KS statistic can then be written as follows,
with a few approximations:
\begin{align*}
E_{G}\left[\text{KS}\left(F_{1}\left(G\right),F_{c}\right)\right] & =\sum_{g}P\left(G=g\right)\text{KS}\left(F_{1}\left(g\right),F_{c}\right)\\
 & =\sum_{l}\sum_{g:\left|E_{g}\right|=l}P\left(G=g\right)\text{KS}\left(F_{1}\left(g\right),F_{c}\right)\\
 & \approx\sum_{l}\sum_{g:\left|E_{g}\right|=l}P\left(G=g\right)\text{KS}\left(\tilde{F_{1}}\left(l\right),F_{c}\right)\\
 & =\sum_{l}\text{KS}\left(\tilde{F_{1}}\left(l\right),F_{c}\right)\sum_{g:\left|E_{g}\right|=l}P\left(G=g\right)\\
 & =\sum_{l}\text{KS}\left(\tilde{F_{1}}\left(l\right),F_{c}\right)P\left(\left|E_{G}\right|=l\right)
\end{align*}
The summation in the first line is over all the possible realizations,
indexed by $g$, of a network $G$ generated by the $G\left(n,p\right)$
model. Assuming the proportion of nodes in the subsample is $\alpha$,
then such an induced subgraph of a network generated by the $G\left(n,p\right)$
model should be $G\left(\alpha n,p\right)$, since each of the possible
edges of the induced subgraph are still independent and have a probability
of $p$ to exist. Thus, $F_{c}$ would still be binomial, namely $B\left(C\left(\alpha n,2\right),p\right)$,
and remains constant. On the other hand, $F_{1}$ depends on $g$
and is indicated as such. On the second line, a new index $l$, the
number of edges in $G$, is introduced, with a nested summation for
all $g$ such that its edge set $E_{g}$ has cardinality $l$. Next,
we approximate $F_{1}\left(g\right)$ for $g$ such that $\left|E_{g}\right|=l$
with $\tilde{F_{1}}\left(l\right)$, which only depends on $l$. Finally,
we can write the nested summation on the fourth line as $P\left(\left|E_{G}\right|=l\right)$,
the probability for $G$ to have $l$ edges.

Let $p_{l}=l/C\left(n,2\right)$, then conditional on $l$, a randomly
selected dyad (node pair) from the induced subgraph is an edge with
probability $p_{l}$. Thus, one reasonable form for $\tilde{F_{1}}\left(l\right)$
would be $B\left(C\left(\alpha n,2\right),p_{l}\right)$. We found
this approximation to be accurate only when $\alpha$ is sufficiently
small ($<0.3$). For larger values of $\alpha$, this approximation ignores
the increasing effect of correlation between different subsamplues due
to increasing number of shared dyads,
leading to underdispersion when using the $B\left(C\left(\alpha n,2\right),p_{l}\right)$
approximation. To correct for the correlation, the covariance between two subsamples
can be derived exactly, allowing for improved approximation of $\tilde{F_{1}}\left(l\right)$.
Let $EC_{1}$ and $EC_{2}$ represent the node count from two different
subsamples of $m$ nodes from an ER graph of $n$ nodes such that
$\left|E_{g}\right|=l$. Then the covariance of $EC_{1}$ and $EC_{2}$
can be written as:
\begin{align*}
\text{cov}\left(EC_{1},EC_{2}\right) & =E\left[EC_{1}\times EC_{2}\right]-E\left[EC_{1}\right]E\left[EC_{2}\right]
\end{align*}
The form for $E\left[EC_{i}\right]=C\left(m,2\right)\times p_{l}$
is simple, but the first term is more involved. Let $m^{*}=C\left(m,2\right)$,
$o^{*}=C\left(o,2\right)$, $e_{j}^{i}$ be the edge indicator for
the $j$th dyad in the $i$th subsample, where $\mathbb{O}$ s.t.
$\left|\mathbb{O}\right|=o$ is the set of nodes that overlap between
the two subsamples. Then the second term can be written as:
\begin{align*}
E\left[EC_{1}\times EC_{2}\right] & =\sum_{o}E\left[\left.\left(\sum_{j=1}^{m^{*}}e_{j}^{1}\right)\left(\sum_{j=1}^{m^{*}}e_{j}^{2}\right)\right|\left|\mathbb{O}\right|=o\right]\times P\left(\left|\mathbb{O}\right|=o\right)\\
 & =\sum_{o}A_{o}\times B_{o}\\
A_{o} & =o^{*}\times p_{l}+2\times C\left(o^{*},2\right)p_{l}^{2}+2\left(m^{*}-o^{*}\right)o^{*}p_{l}^{2}+\left(m^{*}-o^{*}\right)^{2}p_{l}^{2}\\
B_{o} & \sim C\left(n,2m-o\right)\times C\left(2m-o,o\right)\times C\left(2m-2o,m-o\right)
\end{align*}
The detailed derivation for $A_{o}$ and $B_{o}$ are shown in the
appendix. We also show in the appendix how to generalize the results
to other models under dyadic independence, as well as for without dyadic
independence.

Two pieces are needed to compute the last line of the expression
for $E_{G}\left[\text{KS}\left(F_{1}\left(G\right),F_{c}\right)\right]$.
$P\left(\left|E_{G}\right|=l\right)$ is simple to compute since $G$
is $G\left(n,p\right)$, thus $\left|E_{G}\right|$ is distributed
according to $B\left(C\left(n,2\right),p\right)$. $\text{KS}\left(\tilde{F_{1}}\left(l\right),F_{c}\right)$
is less straightforward, but can be approximated using normal approximations.
$F_{c}$ can be approximated with a normal distribution with the corresponding
binomial mean and variance, $N\left(C\left(\alpha n,2\right)p,C\left(\alpha n,2\right)p\left(1-p\right)\right)$.
For $\tilde{F_{1}}\left(l\right)$, the naive approximation that ignores
correlation is similarly $N\left(C\left(\alpha n,2\right)p_{l},C\left(\alpha n,2\right)p_{l}\left(1-p_{l}\right)\right)$.
However, as stated above, this approximation is inaccurate for larger
values of $\alpha$. We found that an approximation with a normal
distribution with mean $C\left(\alpha n,2\right)p_{l}$ and variance
$E\left[EC_{1}^{2}\right]-E\left[EC_{1}\right]E\left[EC_{2}\right]-\text{cov}\left(EC_{1},EC_{2}\right)$
yields a much closer approximation. Assume that $p\ne p_{l}$, then
it can be easily verified that the maximal difference between the
two normal CDFs occurs at $x_{l}$, the point where the two normal
density functions are equal. Thus, for a particular value of $l$,
$\text{KS}\left(\tilde{F_{1}}\left(l\right),F_{c}\right)$ can be
approximated by the absolute value of the difference between the two
normal CDFs evaluated at $x_{l}$.

\begin{center}
{\scriptsize{}}%
\begin{tabular}{|c|c|c|c|c|c|c|c|c|c|c|c|}
\hline 
{\scriptsize{}$\alpha$} & {\scriptsize{}0.05} & {\scriptsize{}0.1} & {\scriptsize{}0.15} & {\scriptsize{}0.2} & {\scriptsize{}0.25} & {\scriptsize{}0.3} & {\scriptsize{}0.5} & {\scriptsize{}0.6} & {\scriptsize{}0.7} & {\scriptsize{}0.8} & {\scriptsize{}0.9}\tabularnewline
\hline 
{\scriptsize{}$E_{G}\left[\text{KS}\left(F_{1}\left(G\right),F_{c}\right)\right]$
(naive)} & {\scriptsize{}0.0158} & {\scriptsize{}0.0317} & {\scriptsize{}0.0475} & {\scriptsize{}0.0633} & {\scriptsize{}0.0790} & {\scriptsize{}0.0947} & {\scriptsize{}0.1559} & {\scriptsize{}0.1855} & {\scriptsize{}0.2143} & {\scriptsize{}0.2422} & {\scriptsize{}0.2692}\tabularnewline
\hline 
{\scriptsize{}$E_{G}\left[\text{KS}\left(F_{1}\left(G\right),F_{c}\right)\right]$
(improved)} & {\scriptsize{}0.0158} & {\scriptsize{}0.0319} & {\scriptsize{}0.0482} & {\scriptsize{}0.0650} & {\scriptsize{}0.0821} & {\scriptsize{}0.0999} & {\scriptsize{}0.1792} & {\scriptsize{}0.2265} & {\scriptsize{}0.2824} & {\scriptsize{}0.3525} & {\scriptsize{}0.4517}\tabularnewline
\hline 
{\scriptsize{}$\hat{E}_{G}\left[\text{KS}\left(F_{1}\left(G\right),F_{c}\right)\right]$} & {\scriptsize{}0.0228} & {\scriptsize{}0.0383} & {\scriptsize{}0.0518} & {\scriptsize{}0.0690} & {\scriptsize{}0.0853} & {\scriptsize{}0.1033} & {\scriptsize{}0.1815} & {\scriptsize{}0.2284} & {\scriptsize{}0.2835} & {\scriptsize{}0.3532} & {\scriptsize{}0.4511}\tabularnewline
\hline 
\end{tabular}{\scriptsize\par}
\par\end{center}
\begin{description}
\item [{Table}] 1: Theoretical approximation and empirical estimate of
$E_{G}\left[\text{KS}\left(F_{1}\left(G\right),F_{c}\right)\right]$
at various values of $\alpha$, the proportion of nodes in each subsample.
\end{description}

Next, we examined the relationship between $\alpha$ and $E_{G}\left[\text{KS}\left(F_{1}\left(G\right),F_{c}\right)\right]$
in a numerical example. We computed $E_{G}\left[\text{KS}\left(F_{1}\left(G\right),F_{c}\right)\right]$,
with the above approximations, for $n=1000$, $p=0.2$, and $\alpha\in\{$0.05, 0.1, 0.15, 0.2, 0.25, 0.3, 0.5, 0.6, 0.7, 0.8, 0.9\}.
In addition, we empirically estimated $E_{G}\left[\text{KS}\left(F_{1}\left(G\right),F_{c}\right)\right]$
for each value of $\alpha$ via simulation, where $F_{1}\left(g\right)$
is estimated from 10000 subsamples of each of 250 independent draws
from $G\left(1000,0.2\right)$ and $F_{c}$ is estimated from single
subsamples of 10000 independent draws from $G\left(1000,0.2\right)$.
The results are summarized in \textbf{Table} 1.

Clearly, $E_{G}\left[\text{KS}\left(F_{1}\left(G\right),F_{c}\right)\right]$
increases with $\alpha$, although not greatly in the lower range
of values of $\alpha$ explored. The naive approximation matches the
empirical results closely until about $\alpha=0.3$, but is very inaccurate
for larger values of $\alpha$. The improved approximation matches
the empirical results closely for all values of $\alpha$ and dominates
the naive approximation for all values of $\alpha$ examined. The
discrepancy between $F_{1}$ and $F_{c}$ does increase with $\alpha$,
but remains small for reasonably small values of $\alpha$. The improved
approximation seems to adhere more closely to empirical results for
larger values of $\alpha$ where more nodes are sampled. This is expected
since the normal approximation for the binomial distribution improves
with larger number of trials. Although this is merely a toy example
and the results are by no means general, they do suggest to keep the
portion of nodes in the subsample low ($<30\%$ in this example) as long as sufficiently many
features of the models can be retained. In addition, this is a cautionary
tale about the care needed in choosing the proportion of nodes sampled,
since even under dyadic independence, the difference between $F_{1}$
and $F_{c}$ can be noticeably larger than an intuitive approximation
for certain values of $\alpha$.

\section{Proposed Usage}

There are a variety of statistical procedures that can take advantage
of this sampling scheme, with a few of them detailed below. Before
proposing the general framework for a few typical statistical procedures
via the bootstrap subsampling procedure, we define the following notation
for the rest of the section. The observed network will be referred
to as $G_{o}$ with $B_{o}$ subsamples and corresponding induced
subgraphs $\tilde{G}_{o}^{\left(1\right)}\ldots\tilde{G}_{o}^{\left(B_{o}\right)}$.
The draws from candidate model $M_{c}$ will be referred to as $G_{M}^{1}\ldots G_{M}^{B_{M}}$
with corresponding subsample induced subgraphs $\tilde{G}_{M}^{\left(1\right)}\ldots\tilde{G}_{M}^{\left(B_{M}\right)}$.
Given a set of network statistics $S$ chosen for model selection or assessing
goodness of fit, the set computed from $\tilde{G}_{o}^{\left(1\right)}\ldots\tilde{G}_{o}^{\left(B_{o}\right)}$
will be referred to as $\tilde{S}_{o}^{\left(1\right)}\ldots\tilde{S}_{o}^{\left(B_{o}\right)}$,
while those computed from $\tilde{G}_{M}^{\left(1\right)}\ldots\tilde{G}_{M}^{\left(B_{M}\right)}$
will be referred to as $\tilde{S}_{M}^{\left(1\right)}\ldots\tilde{S}_{M}^{\left(B_{M}\right)}$.
Note that $B_{o}$ and $B_{M}$ need not be equal.

\subsection{Model Selection}

Suppose the goal is to select between candidate models $M_{1}\ldots M_{c}$
for $G_{o}$. Given a set of statistics $S$ to base the model selection
on, one needs to compute $\tilde{S}_{M_{i}}^{\left(1\right)}\ldots\tilde{S}_{M_{i}}^{\left(B_{M}\right)}$
from $\tilde{G}_{M_{i}}^{\left(1\right)}\ldots\tilde{G}_{M_{i}}^{\left(B_{M}\right)}$
for $i=1\ldots c$. These collections of statistics along with the
model indices of each draw form the training data and are the basis
for the model selection procedure. The selection of $S$ is flexible
and should be chosen to prioritize the aspects of the network where
similarity to the observed network is most paramount. The training
data can be used to train any learning algorithm for prediction of
the model index. Examples include random forest, support vector machine,
or even ensemble learning algorithms like Super Learner \citep{SLbookPolley2011super,SLvan2007super,chen2018SLms}.
Lastly, the trained algorithm can be evaluated at each of $\tilde{S}_{o}^{\left(1\right)}\ldots\tilde{S}_{o}^{\left(B_{o}\right)}$
to give selected model $\hat{M}_{1}\ldots\hat{M}_{B_{o}}$, with majority
rule deciding the final selected model.

\begin{description}
\item [{Algorithm}] I: Steps for the model selection with the bootstrap
subsampling procedure.
\end{description}
\begin{enumerate}
\item Draw subsamples $\tilde{G}_{o}^{\left(1\right)}\ldots\tilde{G}_{o}^{\left(B_{o}\right)}$
from $G_{o}$
\item Draw subsamples $\tilde{G}_{M_{i}}^{\left(1\right)}\ldots\tilde{G}_{M_{i}}^{\left(B_{M}\right)}$
from each candidate model $i=1\ldots c$
\item Compute statistics for model selection for $\tilde{G}_{o}^{\left(1\right)}\ldots\tilde{G}_{o}^{\left(B_{o}\right)}$
and $\tilde{G}_{M_{i}}^{\left(1\right)}\ldots\tilde{G}_{M_{i}}^{\left(B_{M}\right)}$
for each $i = 1 \ldots c$
\item Form training data based on each of $\tilde{S}_{M_{i}}^{\left(1\right)}\ldots\tilde{S}_{M_{i}}^{\left(B_{M}\right)}$
along with model index $i$
\item Train learning algorithm based on training data where the predictors
are the network statistics and the outcome is the model index $i$
\item Evaluate trained algorithm on $\tilde{S}_{o}^{\left(1\right)}\ldots\tilde{S}_{o}^{\left(B_{o}\right)}$
and select the model based on plurality rule
\end{enumerate}

One distinct advantage of the model selection through this bootstrap
subsampling procedure is that it gives inherent evidence about uncertainty
or confidence in the selected model as well as other candidate models.
The proportion of $\tilde{G}_{o}^{\left(1\right)}\ldots\tilde{G}_{o}^{\left(B_{o}\right)}$
that are assigned to each model can be seen as evidence in favor
of each candidate model, while the proportion of subsamples assigned
the model that forms the majority can be seen as confidence in the
selected model. With algorithms like random forest, where the decision
is based on majority rule as well, this does not add anything new.
But with others, such as support vector machine or the Super Learner
that are not based on majority rule, this approach offers a way to
quantify uncertainty without the need to alter the learning algorithm
in any way.

\subsection{Goodness of Fit}

To assess the goodness of fit for candidate models $M_{1}\ldots M_{c}$,
the procedure is similar to that of model selection. For a set of
statistics $S$ for assessing goodness of fit, one computes $\tilde{S}_{o}^{\left(1\right)}\ldots\tilde{S}_{o}^{\left(B_{o}\right)}$
from $\tilde{G}_{o}^{\left(1\right)}\ldots\tilde{G}_{o}^{\left(B_{o}\right)}$
and $\tilde{S}_{M_{i}}^{\left(1\right)}\ldots\tilde{S}_{M_{i}}^{\left(B_{M}\right)}$
from $\tilde{G}_{M_{i}}^{\left(1\right)}\ldots\tilde{G}_{M_{i}}^{\left(B_{M}\right)}$
for $i=1\ldots c$. Rather than training a learning algorithm based
on $\tilde{S}_{M_{i}}^{\left(1\right)}\ldots\tilde{S}_{M_{i}}^{\left(B_{M}\right)}$
as in model selection, $\tilde{S}_{o}^{\left(1\right)}\ldots\tilde{S}_{o}^{\left(B_{o}\right)}$
can be directly compared against $\tilde{S}_{M_{i}}^{\left(1\right)}\ldots\tilde{S}_{M_{i}}^{\left(B_{M}\right)}$
for each $i$ to assess fit. As mentioned above, this comparison between
the distribution of $\tilde{S}_{o}^{\left(1\right)}\ldots\tilde{S}_{o}^{\left(B_{o}\right)}$
and any set of $\tilde{S}_{M_{i}}^{\left(1\right)}\ldots\tilde{S}_{M_{i}}^{\left(B_{M}\right)}$
can be done in terms of location, spread, shape, or other aspects
of the distribution. This can be done visually by comparing the histograms
of the two resampling distributions, but also numerically by comparing
the mean and variance of the two.

\begin{description}
\item [{Algorithm}] II: Steps for assessing goodness of fit with the bootstrap
subsampling procedure.
\end{description}
\begin{enumerate}
\item Draw subsamples $\tilde{G}_{o}^{\left(1\right)}\ldots\tilde{G}_{o}^{\left(B_{o}\right)}$
from $G_{o}$
\item Draw subsamples $\tilde{G}_{M_{i}}^{\left(1\right)}\ldots\tilde{G}_{M_{i}}^{\left(B_{M}\right)}$
from each candidate model $i=1\ldots c$
\item Compute $\tilde{S}_{o}^{\left(1\right)}\ldots\tilde{S}_{o}^{\left(B_{o}\right)}$
and $\tilde{S}_{M_{i}}^{\left(1\right)}\ldots\tilde{S}_{M_{i}}^{\left(B_{M}\right)}$
from $\tilde{G}_{o}^{\left(1\right)}\ldots\tilde{G}_{o}^{\left(B_{o}\right)}$
and $\tilde{G}_{M_{i}}^{\left(1\right)}\ldots\tilde{G}_{M_{i}}^{\left(B_{M}\right)}$,
respectively
\item Assess fit by comparing $\tilde{S}_{o}^{\left(1\right)}\ldots\tilde{S}_{o}^{\left(B_{o}\right)}$
and $\tilde{S}_{M_{i}}^{\left(1\right)}\ldots\tilde{S}_{M_{i}}^{\left(B_{M}\right)}$
\end{enumerate}

Assessment based on any one of these aspects may however lead to conflicting
results, i.e., different models having the best fit depending on which
aspect the comparison is based on, and it might be desirable to make
comparisons through a more holistic measure. One solution to this
is to compute a distance measure, such as the KS statistic or the
Kullback-Leibler divergence, between $\tilde{S}_{o}^{\left(1\right)}\ldots\tilde{S}_{o}^{\left(B_{o}\right)}$
and $\tilde{S}_{M_{i}}^{\left(1\right)}\ldots\tilde{S}_{M_{i}}^{\left(B_{M}\right)}$
to quantify the fit of model $i$. This gives a single statistic that
takes the entire distribution into account to quantify and to categorically
order the fit of each candidate model. The KS test statistic and Kullback-Leibler
divergence are typically computed in one dimension and can be used
to compare the fit for each statistic individually as is. Instead,
should one wish to make a comparison based on all statistics $S$
at the same time, one can look to use generalizations of these statistics
\citep{peacock1983two,fasano1987multidimensional,justel1997multivariate}.

\subsection{Comparison of Multiple Networks}

If multiple networks are observed instead of a single network, and the goal is to assess
how similar they are, then one can do so by building a resampling
distribution from multiple networks. For the case of two observed
networks with a set of statistics $S$ for comparison and observed
networks $G_{o1}$ and $G_{o2}$, one can compute $\tilde{S}_{o1}^{\left(1\right)}\ldots\tilde{S}_{o1}^{\left(B_{o1}\right)}$
and $\tilde{S}_{o2}^{\left(1\right)}\ldots\tilde{S}_{o2}^{\left(B_{o2}\right)}$
from subsamples $\tilde{G}_{o1}^{\left(1\right)}\ldots\tilde{G}_{o1}^{\left(B_{o1}\right)}$
and $\tilde{G}_{o2}^{\left(1\right)}\ldots\tilde{G}_{o2}^{\left(B_{o2}\right)}$.
The comparison of the two is based on $\tilde{S}_{o1}^{\left(1\right)}\ldots\tilde{S}_{o1}^{\left(B_{o1}\right)}$
and $\tilde{S}_{o2}^{\left(1\right)}\ldots\tilde{S}_{o2}^{\left(B_{o2}\right)}$,
and one can proceed essentially the same way as with goodness of fit
by comparing different aspects of the two distributions, but with
$\tilde{S}_{o1}^{\left(1\right)}\ldots\tilde{S}_{o1}^{\left(B_{o1}\right)}$
and $\tilde{S}_{o2}^{\left(1\right)}\ldots\tilde{S}_{o2}^{\left(B_{o2}\right)}$
in place of $\tilde{S}_{o}^{\left(1\right)}\ldots\tilde{S}_{o}^{\left(B_{o}\right)}$
and $\tilde{S}_{M_{i}}^{\left(1\right)}\ldots\tilde{S}_{M_{i}}^{\left(B_{M}\right)}$.
Should there be more than two observed networks for comparison, then
the distance measure statistics can once again be used to quantify
all pairwise relative similarities between the observed networks.

\section{Simulation and Data Examples}

We use a few simulation studies as well as data from an empirical
network to illustrate the use of the bootstrap subsampling procedure
in some of the scenarios described in the previous section.

\subsection{Model Selection}

The simulation studies conducted for model selection consider instances
of a variation on the afformentioned $G\left(n,m\right)$ model we
introduced \citep{chen2018SLms}. This variation generates random
graphs with $n$ nodes and $m$ edges just as the $G\left(n,m\right)$
model with each edge being added one at a time. At each step in network
generation, a pair of unconnected nodes are selected at random, and
the probability for adding an edge between the two is determined based
on the number of triangles it would close, then the edge is added
with the given probability. This is repeated until there are $m$
edges in the network. If the probability for adding an edge is fixed,
then this is the $G\left(n,m\right)$ model. Instead, we start with
a base probability $p_{0}$ to add the edge. Should the edge close
at least one triangle, the probability increases by $p_{1}$. Finally,
should multiple triangles be closed by the edge, then the probability
further increases by $p_{2}$ for each additional triangle closed.

In the simulation, we select between two instances of this model,
both having $p_{0}=0.3$ and $p_{1}=0.1$. The difference comes in
$p_{2}$, with $p_{2}=0$ for model 1, while $p_{2}$ varies over
0.05, 0.03, 0.01, 0.005 for model 2. For given choices of $n$ and
$m$, as $p_{2}$ decreases and gets closer to 0, the difference between
the two models become more difficult to detect. The generated networks
consist of 100 nodes with edge count varying over 100, 500, 1000,
2000. For a given set of parameter values, the difference between
the two models should be easier to detect as edge count increases,
since attenuation from the difference in $p_{2}$ has more opportunities
to manifest itself. The training data consists of a single subsample
of 80 nodes for each of 10000 draws from each model ($\tilde{G}_{M_{i}}^{\left(1\right)}\ldots\tilde{G}_{M_{i}}^{\left(10000\right)}$).
The test data consists of 1000 draws from each model ($G_{o}$), while
the model selection is based on 100 subsamples of 80 nodes from each
draw ($\tilde{G}_{o}^{\left(1\right)}\ldots\tilde{G}_{o}^{\left(100\right)}$).

The model selection is through the Super Learner (see citations in
section 5.2 for details), with support vector machine, random forest,
and $k$-nearest neighbors as candidate algorithms, and average clustering
coefficient, triangle count, as well as the three quartiles of the
degree distribution as predictors. These statistics were chosen as
predictors since the difference in $p_{2}$ directly affects formation
of triangles, while the other statistics are influenced strongly by
triangles. For each of the 100 $\tilde{G}_{o}^{\left(b_{o}\right)}$
for a particular testing network $G_{o}$, the Super Learner will give
a score between 0 and 1 for predicting the model class of $\tilde{G}_{o}^{\left(b_{o}\right)}$,
with score \textless 0.5 assigned model 1 and score \textgreater 0.5
assigned model 2. The selected model is the model assigned to more
$\tilde{G}_{o}^{\left(b_{o}\right)}$s, i.e., the majority of model
assignment.

The results of the simulation are summarized in \textbf{Figure} 5
and \textbf{Table} 2. \textbf{Table} 2 contains the proportion of
test networks whose model was correctly classified by the Super Learner
at each combination of $p_{2}$ and edge count. Unsurprisingly, the
proportion decreases as $p_{2}$ decreases for a fixed edge count,
and increases as edge count increases for a fixed $p_{2}$. \textbf{Figure} 5
shows the histogram of the confidence for the correct model. When
model 1 is the true model of the test network, this is the proportion
of the 100 subsamples that were assigned model 1, and vice versa.
When the proportion of correctly classified models is around 0.5,
i.e., as good as a random guess, the confidence is symmetric and centered
close to 0.5. When the proportion is higher than 0.5, the distribution
of the confidence is shifted to the right, meaning that the two models
are easier to tell apart. In addition, the more right skewed the
histograms, the more confidence in the correct model. The red vertical
line indicates the median, which also moves to the right as the proportion
increases and as the confidence becomes more right skewed. This behavior
indicates that the confidence for the selected model from the bootstrap
subsampling procedure quantifies well the degree of uncertainty in
the selected model.

\begin{center}
\begin{tabular}{|c|c|c|c|c|}
\hline 
 & $p_{2}=0.05$ & 0.03 & 0.01 & 0.005\tabularnewline
\hline 
Edge count = 100 & 0.5015 & 0.5005 & 0.4834 & 0.5100\tabularnewline
\hline 
500 & 0.6092 & 0.5670 & 0.5178 & 0.5076\tabularnewline
\hline 
1000 & 0.9203 & 0.8202 & 0.6249 & 0.5786\tabularnewline
\hline 
2000 & 0.9890 & 0.9740 & 0.8343 & 0.6810\tabularnewline
\hline 
\end{tabular}
\par\end{center}
\begin{description}
\item [{Table}] 2: Proportion of the test networks correctly classified
at each combination of $p_{2}$ and edge count.
\end{description}

\begin{center}
\includegraphics[width=\linewidth]{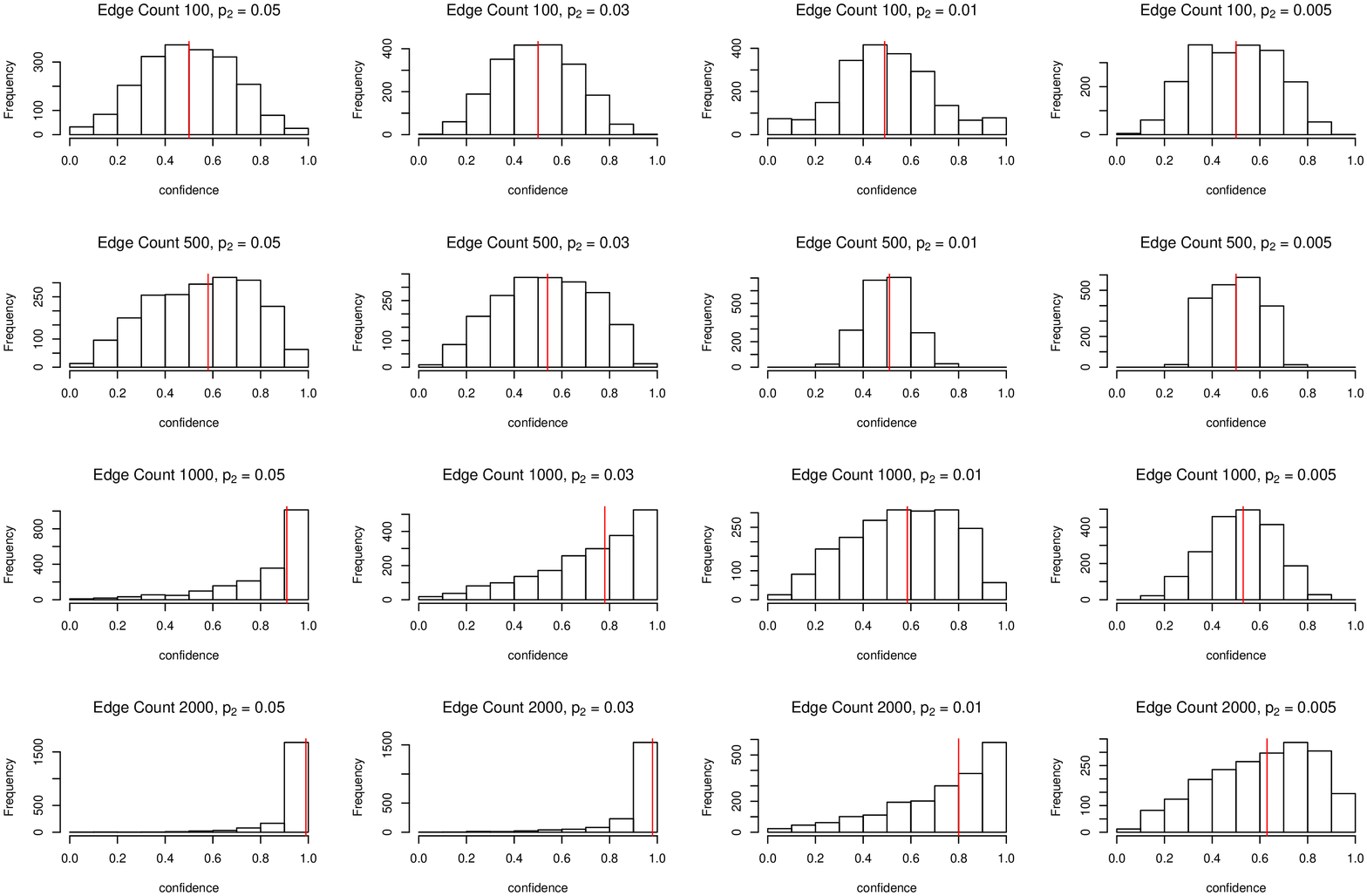}
\par\end{center}
\begin{description}
\item [{Figure}] 5: Histograms of the confidence score (proportion of subsamples
assigned the correct model here rather than the majority) for $p_{2}$
from 0.05, 0.03, 0.01, 0.005, from left to right, and edge count from
100, 500, 1000, 2000, from top to bottom, with the red vertical lines representing
the median.
\end{description}

\subsection{Goodness of Fit}

To display our method for assessment of goodness of fit, we examine
the yeast (\textit{S.cerevisiae}) protein-protein interaction network
data from the database of interacting proteins (DIP) \citep{salwinski2004database}.
This data set has been much examined in the literature, including
via network models. There are two particular publications \citep{hormozdiari2007not,schweiger2011generative}
that fit different duplication divergence models to two different
previous versions of the yeast data set, with differing seed networks.
Here we apply our method to compare the fit of the two different models
on the most recent version of the data.

Both papers use the same duplication divergence model \citep{sole2002model,pastor2003evolving},
which we described as DMR in section 3.2. However, the papers used different
parameter values as well as different seed networks. The fit from
\citet{hormozdiari2007not} has parameter values $p=0.365$ and $r=0.12$,
and the seed network contains 50 nodes. The seed network\footnote{Note that the details for obtaining the seed network from \citet{hormozdiari2007not}
was somewhat incomplete, so this is our interpretation of the description
of their seed network.} was constructed by highly connecting cliques, complete graphs where
an edge exists between every pair of nodes, of 7 nodes and 10 nodes,
then connecting additional nodes to the cliques. To highly connect
the cliques, each possible edge between nodes in different cliques
(70 such edges) was added with probability 0.67. Then, another 33
nodes were attached to randomly chosen nodes from the two cliques.
At each step of the network generation, if a singleton (a node not
connected to any other node) was generated, it was immediately removed
in their model.

On the other hand, the fit from \citet{schweiger2011generative} has
parameter values $p=0.3$ and $r=1.05$. They use a smaller seed network
of 40 nodes, generated with an inverse geometric model. To generate
this seed network, a set of coordinates $\left\{ x_{1}\ldots x_{40}\right\} $
in $\mathbb{R}^{d}$ is generated for each node. Then, each pair of
nodes with distance $\left\Vert x_{i}-x_{j}\right\Vert $ greater
than some threshold $R$ is connected with an edge. Each dimension
of the coordinates is independently generated from the standard normal
distribution $N\left(0,1\right)$. In their fit, the seed network
uses $d=2$ and $R=1.5$. Unlike \citet{hormozdiari2007not}, \citet{schweiger2011generative}
does not remove singletons as they are generated.

Both papers assessed the fit of their model by comparing certain aspects
of the generated network to those of the yeast PPI network. In \citet{hormozdiari2007not},
the fit of their model was assess via $k$-hop reachability, the number
of distinct nodes reachable in $\le k$ edges, the distribution of
particular subraphs, such as triangles and stars, as well as some
measures of centrality. \citet{schweiger2011generative} does so with
the distribution of bicliques, i.e., subgraphs of two disjoint sets
of nodes where every possible edge between the two sets exists. Here,
we assess the fit of both models via our method with the average local
clustering coefficient, triangle count, as well as the degree assortativity.
The local clustering coefficient of a particular node is a measure
of how much does its neighbors resemble a clique. Mathematically,
this is computed as the number of edges between a node's neighbors
divided by the possible number of such edges. We use the average of
the local clustering coefficient over all nodes in the network as
a meassure of local clustering that is also attributable to the network
as a whole. We also consider the number of triangle subgraphs that appear
in the network. Unlike \citet{hormozdiari2007not}, which counts the
total number of various subgraphs together, the count of triangles
alone is a strictly global measure of clustering. Lastly, the degree
assortativity of a network is a measure of how similar are the degrees
of nodes connected by an edge. It is defined as the Pearson correlation
of the degrees of nodes connected by an edge, so positively assorted
networks have more edges between nodes of similar degrees, while negatively
assorted networks have more edges between nodes of dissimilar degrees.

For the analysis, we consider the largest connected component (LCC)
of the PPI network just as in \citet{hormozdiari2007not}. The full
network from the current version of the data contains 5176 nodes and
22977 edges, while the LCC contains 5106 (98.6\%) nodes and 22935 (99.8\%) edges. Networks
drawn from each model contains the same number of nodes as the LCC,
starting from their respective seed networks described above. Subsamples
from the PPI network as well as networks drawn from each model contain
1550 nodes, roughly corresponding to 30\%. This was the largest portion
considered in section 4.

The results of the data analysis are summarized in \textbf{Figure}
6. From the figures, it's clear that the ordering of the fit of both
models differ based on the statistic of comparison. For clustering
coefficient, both models fit equally poorly, as the resampling distribution
of both models and that of the PPI network have no overlap at all.
The KS statistic between the resampling distribution of the PPI network
and that of each model are both 1, indicating very poor fit. The distance
between the location of both models' resampling distribution and that
of the PPI network are very similar, so existing methods that assess
goodness of fit based on point estimates only likely would arrive
at the same conclusion. For triangle count, the model of \citet{schweiger2011generative}
seems to fit better as its resampling distribution's spread has a
much bigger overlap with that of the PPI network. The KS statistic
for the model of \citet{schweiger2011generative} (0.6778) is also
much smaller than that of \citet{hormozdiari2007not} (0.9018). However,
unlike clustering coefficient, the distance between the location of
both models' resampling distribution and that of the PPI network are
rather similar, so existing methods likely would have concluded that
the fit of both models are similar in this regard. Lastly, for degree
assortativity, the model of \citet{hormozdiari2007not} fits much
better as all of the spread of its resampling distribution overlaps
with that of the PPI network, and most of its spread is negative just
as the PPI network, indicating negative degree assortativity. On the
other hand, the resampling distribution of the model of \citet{schweiger2011generative}
is entirely positive and has little overlap with that of the PPI network.
The KS statistic tells the same story, with 0.4373 for \citet{hormozdiari2007not}
and 0.9782 for \citet{schweiger2011generative}. The distance of the
location of the two models' resampling distributions to that of the
PPI network are very distinct, so exisiting methods would likely reach
the same conclusion. We can see situations in this data set where
existing methods and our method would reach the same conclusion, but
also where the two would reach different conclusions due to the additional
layer of information encoded in the resampling distributions.

In addition in \textbf{Figure} 6, we plot the subsamples from two
individual networks drawn from each model against the subsamples from
independent networks drawn from each model. For each statistic, the
spread and location of the two types of subsamples are similar. This
is likely due to the rather large seeds (50 and 40 nodes respectively)
both models use as well as the rather small portion of nodes in each
subsample ($\sim 30\%$), reflecting observations from sections
3 and 4.

\begin{center}
\includegraphics[width=\linewidth]{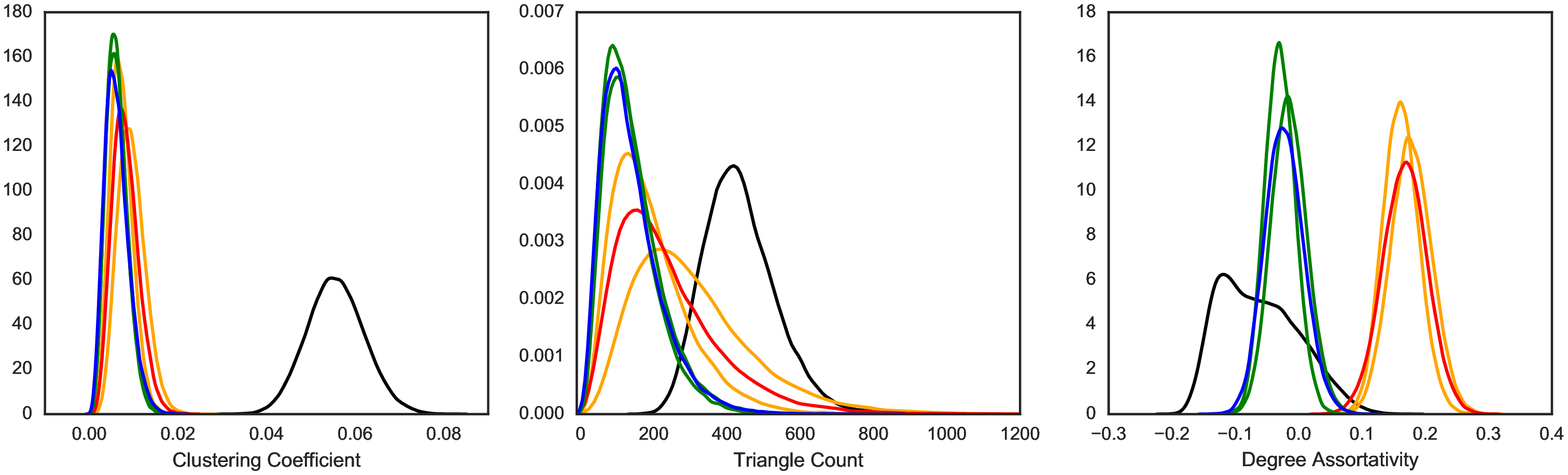}
\par\end{center}
\begin{description}
\item [{Figure}] 6: The resampling distribution of clustering coefficient,
triangle count, and degree assortativity (left to right) from independent
draws from the two model fits (blue for \citet{hormozdiari2007not}
and red is for \citet{schweiger2011generative}) as well as the PPI
network (black). In addition, there are two resampling distributions
from a single draw from each of the two model fits (green for \citet{hormozdiari2007not}
and orange is for \citet{schweiger2011generative}).
\end{description}

\section{Discussion}

Network models continue to expand the amount of correlation they can incorporate
and are able to model increasingly complex dependencies that can arise
in network data. Yet this very dependency poses a statistical challenge,
especially in the case of a single observed network. We propose a
bootstrap subsampling procedure as a basis for statistical procedures
in this setting that is based on a flexible resampling distribution
built from the single observed network.

Given any statistic of interest, its corresponding resampling distribution
can be compared against its analog from a null/candidate model based
on any attribute of their distributions, including, but not limited
to, location, spread, shape, measures of mean, as well as distances.
In comparison, existing methods in this setting typically rely on
the point estimate from the observed network, which leads to a more
limited comparison. As seen in our data example, this additional layer
of information can sometimes lead to a different conclusion than existing
methods. In addition, the distance between the resampling distributions
leads to a single holistic measure for comparison as well as ordering
of different network models.

The flexibility in our approach is not limited to what one can do
with these resampling distributions, but also the type of subsampling
used to generate them. Although in the simulation and data example,
the subsamples are simply random samples of the nodes of the network,
they need not always be. In fact, any method of subsampling is valid
as long as it is applied to both the observed data and the null/candidate
model. Thus, it can be tailored to any needs of the investigator,
such as statistical or computational considerations. The method of
subsampling can be also used as a sensitivity analysis to see whether
the results of the analysis remain unchanged under different methods
of subsampling. This consideration for different methods of subsampling
motivates the most immediate step for future work as it begs the question
whether they can lead to performance gains. Perhaps certain types
of subsampling can outperform others given the method of sampling
used to obtain the observed data.

\section*{Acknowledgements}
S.C. and J.P.O. are both supported by NIH 1DP2MH103909-01.
In addition, S.C. is supported by NIH 5U01HG009088-02 and U54GM088558-09;
J.P.O is supported by NIH 5R37AI051164-12, 1R01AI112339-01, and U54GM088558-06.

\section*{Appendix}

Following the uniform nodewise subsampling from the paper for the
ER model. To better approximate the resampling distribution, we need
to estimate the covariance between different subsamples of the same
size from the same ER graph.

Say the full ER graph has $n$ nodes, and each subsample contains
$m$ nodes. Let $EC_{1}$ and $EC_{2}$ represent the node count from
two different subsamples of $m$ nodes from an ER graph of $n$ nodes
containing $p_{l}\times C\left(n,2\right)$ edges:
\begin{align*}
\text{cov}\left(EC_{1},EC_{2}\right) & =E\left[EC_{1}\times EC_{2}\right]-E\left[EC_{1}\right]E\left[EC_{2}\right]
\end{align*}
$E\left[EC_{i}\right]=C\left(m,2\right)\times p_{l}$, so we need
to focus on the first term. Let $m^{*}=C\left(m,2\right)$, $o^{*}=C\left(o,2\right)$,
$e_{j}^{i}$ be the edge indicator for the $j$th dyad in the $i$th
subsample, and $\mathbb{O}$ be the set of nodes that overlap between
the two subsamples:
\begin{align*}
E\left[EC_{1}\times EC_{2}\right] & =\sum_{o}E\left[\left.\left(\sum_{j=1}^{m^{*}}e_{j}^{1}\right)\left(\sum_{j=1}^{m^{*}}e_{j}^{2}\right)\right|\left|\mathbb{O}\right|=o\right]\times P\left(\left|\mathbb{O}\right|=o\right)=\sum_{o}A_{o}\times B_{o}
\end{align*}

We assess the two terms separately, say the first $o^{*}$ dyads are
from the nodes that overlap:
\begin{align*}
A_{o} & =E\left[\left.\left(\sum_{j=1}^{m^{*}}e_{j}^{1}\right)\left(\sum_{k=1}^{m^{*}}e_{k}^{2}\right)\right|\left|\mathbb{O}\right|=o\right]\\
 & =E\left[\left.\left(\sum_{j=1}^{o^{*}}e_{j}^{1}+\sum_{j=o^{*}+1}^{m^{*}}e_{j}^{1}\right)\left(\sum_{k=1}^{o^{*}}e_{k}^{2}+\sum_{k=o^{*}+1}^{m^{*}}e_{k}^{2}\right)\right|\left|\mathbb{O}\right|=o\right]\\
 & =E\left[\left(\sum_{j=1}^{o^{*}}e_{j}^{1}\right)\left(\sum_{k=1}^{o^{*}}e_{k}^{2}\right)+\left(\sum_{j=1}^{o^{*}}e_{j}^{1}\right)\left(\sum_{k=o^{*}+1}^{m^{*}}e_{k}^{2}\right)\right.\\
 & \left.\left.+\left(\sum_{j=o^{*}+1}^{m^{*}}e_{j}^{1}\right)\left(\sum_{k=1}^{o^{*}}e_{k}^{2}\right)+\left(\sum_{j=o^{*}+1}^{m^{*}}e_{j}^{1}\right)\left(\sum_{k=o^{*}+1}^{m^{*}}e_{k}^{2}\right)\right|\left|\mathbb{O}\right|=o\right]\\
 & =\sum_{j=1}^{o^{*}}\sum_{k=1}^{o^{*}}E\left[e_{j}^{1}e_{k}^{2}\right]+\sum_{j=1}^{o^{*}}\sum_{k=o^{*}+1}^{m^{*}}E\left[e_{j}^{1}e_{k}^{2}\right]+\sum_{j=o^{*}+1}^{m^{*}}\sum_{k=1}^{o^{*}}E\left[e_{j}^{1}e_{k}^{2}\right]+\sum_{j=o^{*}+1}^{m^{*}}\sum_{k=o^{*}+1}^{m^{*}}E\left[e_{j}^{1}e_{k}^{2}\right]\\
 & =\sum_{j=1}^{o^{*}}E\left[\left(e_{j}^{1}\right)^{2}\right]+\sum_{j\ne k\in\left\{ 1\ldots o^{*}\right\} }E\left[e_{j}^{1}\right]E\left[e_{k}^{2}\right]\\
 & +\sum_{j=1}^{o^{*}}\sum_{k=o^{*}+1}^{m^{*}}E\left[e_{j}^{1}\right]E\left[e_{k}^{2}\right]+\sum_{j=o^{*}+1}^{m^{*}}\sum_{k=1}^{o^{*}}E\left[e_{j}^{1}\right]E\left[e_{k}^{2}\right]+\sum_{j=o^{*}+1}^{m^{*}}\sum_{k=o^{*}+1}^{m^{*}}E\left[e_{j}^{1}\right]E\left[e_{k}^{2}\right]\\
 & =\sum_{j=1}^{o^{*}}P\left(e_{j}^{1}=1\right)+2\sum_{j<k\in\left\{ 1\ldots o^{*}\right\} }P\left(e_{j}^{1}=1\right)P\left(e_{k}^{2}=1\right)+\sum_{j=1}^{o^{*}}\sum_{k=o^{*}+1}^{m^{*}}P\left(e_{j}^{1}=1\right)P\left(e_{k}^{2}=1\right)\\
 & +\sum_{j=o^{*}+1}^{m^{*}}\sum_{k=1}^{o^{*}}P\left(e_{j}^{1}=1\right)P\left(e_{k}^{2}=1\right)+\sum_{j=o^{*}+1}^{m^{*}}\sum_{k=o^{*}+1}^{m^{*}}P\left(e_{j}^{1}=1\right)P\left(e_{k}^{2}=1\right)\\
 & =o^{*}\times p_{l}+2\times C\left(o^{*},2\right)p_{l}^{2}+2\left(m^{*}-o^{*}\right)o^{*}p_{l}^{2}+\left(m^{*}-o^{*}\right)^{2}p_{l}^{2}
\end{align*}

In the fourth line, the last three terms are all from products of
distinct dyads, so the expectation of the product can be separated
into product of the expectation. The first term however does contain
some products of the same dyad, and need to be handled differently.
For the first term in the fourth line, since the first $o^{*}$ dyads
are the same in the two subsamples:
\begin{align*}
\sum_{j=1}^{o^{*}}\sum_{k=1}^{o^{*}}E\left[e_{j}^{1}e_{k}^{2}\right] & =\sum_{j=k\in\left\{ 1\ldots o^{*}\right\} }E\left[e_{j}^{1}e_{k}^{2}\right]+\sum_{j\ne k\in\left\{ 1\ldots o^{*}\right\} }E\left[e_{j}^{1}\right]E\left[e_{k}^{2}\right]\\
 & =\sum_{j=1}^{o^{*}}E\left[\left(e_{j}^{1}\right)^{2}\right]+\sum_{j\ne k\in\left\{ 1\ldots o^{*}\right\} }E\left[e_{j}^{1}\right]E\left[e_{k}^{2}\right]
\end{align*}

For $B_{o}$:
\begin{align*}
B_{o} & =P\left(\left|\mathbb{O}\right|=o\right)\\
 & =\frac{\#\text{ of ways to choose two different subsets of }n\text{ elements that have }o\text{ overlapping elements}}{\#\text{ of ways to choose two different subsets of }n\text{ elements}}=\frac{B_{o}^{1}}{B_{o}^{2}}
\end{align*}
We need not compute the denominator, but merely normalize the numerator
for all possible values of $o\in\left\{ \text{max}\left(0,2m-n\right)\ldots m\right\} $.
Note that the union of the two subsets is a set of $2m-o$ elements
\begin{align*}
B_{o}^{1} & =\left(\text{number of ways of choosing }2m-o\text{ elements out of }n\right)\\
 & \times\left(\text{number of ways of choosing the }o\text{ overlapping elements out of }2m-o\right)\\
 & \times\left(\text{number of ways to permute the nonoverlapping }2m-2o\text{ elements between the two subsets}\right)\\
 & =C\left(n,2m-o\right)\times C\left(2m-o,o\right)\times C\left(2m-2o,m-o\right)
\end{align*}

These components allow us to compute $\text{cov}\left(EC_{1},EC_{2}\right)$.
However, to approximate the variance of $EC_{i}$ over different subsamples,
we will use the expectation of the variance estimator. Say we have
taken $B$ subsamples:
\begin{align*}
E\left[\hat{\text{var}}\left(EC_{i}\right)\right] & =E\left[\frac{1}{B}\sum_{i=1}^{B}\left(EC_{i}-\bar{EC}\right)^{2}\right]\\
 & =E\left[\frac{1}{B}\sum_{i=1}^{B}EC_{i}^{2}-\bar{EC}^{2}\right]\\
 & =E\left[EC_{i}^{2}\right]-E\left[\bar{EC}^{2}\right]\\
 & =E\left[EC_{i}^{2}\right]-\frac{1}{B^{2}}E\left[\sum_{i=1}^{B}EC_{i}^{2}+2\sum_{j<k}EC_{j}EC_{k}\right]\\
 & =E\left[EC_{i}^{2}\right]-\frac{1}{B^{2}}\sum_{i=1}^{B}E\left[EC_{i}^{2}\right]-\frac{2}{B^{2}}\sum_{j<k}E\left[EC_{j}EC_{k}\right]\\
 & =\frac{B-1}{B}E\left[EC_{i}^{2}\right]-\frac{2}{B^{2}}C\left(B,2\right)E\left[EC_{j}EC_{k}\right]\\
 & =\frac{B-1}{B}E\left[EC_{i}^{2}\right]-\frac{2}{B^{2}}\frac{B\left(B-1\right)}{2}E\left[EC_{j}EC_{k}\right]\\
 & \approx E\left[EC_{i}^{2}\right]-E\left[EC_{j}EC_{k}\right]\\
 & =E\left[EC_{i}^{2}\right]-E\left[EC_{j}\right]E\left[EC_{k}\right]-\text{cov}\left(EC_{j},EC_{k}\right)\\
 & =\text{var}\left(EC_{i}\right)-\text{cov}\left(EC_{j},EC_{k}\right)
\end{align*}

Results:
\begin{center}
{\scriptsize{}}%
\begin{tabular}{|c|c|c|c|c|c|c|c|c|c|c|c|}
\hline 
{\scriptsize{}$\alpha$} & {\scriptsize{}0.05} & {\scriptsize{}0.1} & {\scriptsize{}0.15} & {\scriptsize{}0.2} & {\scriptsize{}0.25} & {\scriptsize{}0.3} & {\scriptsize{}0.5} & {\scriptsize{}0.6} & {\scriptsize{}0.7} & {\scriptsize{}0.8} & {\scriptsize{}0.9}\tabularnewline
\hline 
{\scriptsize{}$E_{G}\left[\text{KS}\left(F_{1}\left(G\right),F_{c}\right)\right]$
(naive)} & {\scriptsize{}0.0158} & {\scriptsize{}0.0317} & {\scriptsize{}0.0475} & {\scriptsize{}0.0633} & {\scriptsize{}0.0790} & {\scriptsize{}0.0947} & {\scriptsize{}0.1559} & {\scriptsize{}0.1855} & {\scriptsize{}0.2143} & {\scriptsize{}0.2422} & {\scriptsize{}0.2692}\tabularnewline
\hline 
{\scriptsize{}$E_{G}\left[\text{KS}\left(F_{1}\left(G\right),F_{c}\right)\right]$
(improved)} & {\scriptsize{}0.0158} & {\scriptsize{}0.0319} & {\scriptsize{}0.0482} & {\scriptsize{}0.0650} & {\scriptsize{}0.0821} & {\scriptsize{}0.0999} & {\scriptsize{}0.1792} & {\scriptsize{}0.2265} & {\scriptsize{}0.2824} & {\scriptsize{}0.3525} & {\scriptsize{}0.4517}\tabularnewline
\hline 
{\scriptsize{}$\hat{E}_{G}\left[\text{KS}\left(F_{1}\left(G\right),F_{c}\right)\right]$} & {\scriptsize{}0.0228} & {\scriptsize{}0.0383} & {\scriptsize{}0.0518} & {\scriptsize{}0.0690} & {\scriptsize{}0.0853} & {\scriptsize{}0.1033} & {\scriptsize{}0.1815} & {\scriptsize{}0.2284} & {\scriptsize{}0.2835} & {\scriptsize{}0.3532} & {\scriptsize{}0.4511}\tabularnewline
\hline 
\end{tabular}
\par\end{center}{\scriptsize \par}
There is still some discrepancy, but it decreases as the proportion
sampled increases. This is likely due to the normal approximation
being poor when the number of nodes sampled is small.

Regardless of model, the form of $B_{o}$ does not change, given uniform
random sampling. For models under dyadic independence other than ER,
the form of $A_{o}$ changes due to different moments in terms $E\left[\left(e_{j}^{1}\right)^{2}\right]$
and $E\left[e_{j}^{1}\right]E\left[e_{k}^{2}\right]$ in the fifth
line of the above expression for $A_{o}$. For example, with the weighted
ER graph as formulated in \citet{garlaschelli2009weighted}, where each dyad is
assigned weight $W$ with geometric distribution:
\begin{align*}
P\left(W=w\right) & =p^{w}\left(1-p\right)\\
P\left(\text{no edge}\right) & =P\left(W=0\right)=1-p\\
P\left(\text{edge}\right) & =P\left(W>0\right)=p
\end{align*}
Under this formulation:
\begin{align*}
E\left[\left(e_{j}^{1}\right)^{2}\right] & =\frac{p+p^{2}}{\left(1-p\right)^{2}}\\
E\left[e_{j}^{1}\right]=E\left[e_{k}^{2}\right] & =\frac{p}{1-p}
\end{align*}

For models not under dyadic independence, the $E\left[\left(e_{j}^{1}\right)^{2}\right]$
terms on the fifth line are still the second moment of an individual
dyad, but all $E\left[e_{j}^{1}\right]E\left[e_{k}^{2}\right]$ terms
must be replaced with $E\left[e_{j}^{1}e_{k}^{2}\right]$ in order
to properly account for dependence between dyads. The latter can be
obtained from the covariance between dyads as specified by the model.

\newpage{}

\bibliographystyle{abbrvnat}
\bibliography{resampling_paper_ref}

\end{document}